\documentclass[sigconf]{acmart}
\AtBeginDocument{%
  \providecommand\BibTeX{{%
    \normalfont B\kern-0.5em{\scshape i\kern-0.25em b}\kern-0.8em\TeX}}}

\acmConference[MSR 2022]{The 2022 Mining Software Repositories Conference}{May, 2022}{Pittsburgh, PA, USA}
\usepackage{listings}
\usepackage{graphicx}
\usepackage{array}
\newcolumntype{P}[1]{>{\centering\arraybackslash}p{#1}}
\newcolumntype{M}[1]{>{\centering\arraybackslash}m{#1}}
\usepackage{float}
\usepackage{subcaption}
\usepackage{enumitem}
\usepackage{url}

\setcopyright{none}
\pdfoutput=1

\begin{document}

%%
%% The "title" command has an optional parameter,
%% allowing the author to define a "short title" to be used in page headers.
\title{Empirical Standards for Repository Mining}

%%
%% The "author" command and its associated commands are used to define
%% the authors and their affiliations.
%% Of note is the shared affiliation of the first two authors, and the
%% "authornote" and "authornotemark" commands
%% used to denote shared contribution to the research.

% Please add in author names

\author{Preetha Chatterjee}
%\authornote{Both authors contributed equally to this research.}
\email{preetha.chatterjee@drexel.edu}
%\orcid{1234-5678-9012}
%\author{G.K.M. Tobin}
%\authornotemark[1]
%\email{webmaster@marysville-ohio.com}
\affiliation{%
  \institution{Drexel University}
%  \streetaddress{P.O. Box 1212}
    \city{Philadelphia}
  \state{PA}
  \country{USA}
  \postcode{19104}
}

\author{Tushar Sharma}
\email{tushar@dal.ca}
\affiliation{%
  \institution{Dalhousie University}
    \city{Halifax}
  \state{NS}
  \country{Canada}
}

\author{Paul Ralph}
\email{paulralph@dal.ca}
%\orcid{1234-5678-9012}
\affiliation{%
  \institution{Dalhousie University}
%  \streetaddress{P.O. Box 1212}
    \city{Halifax}
  \state{NS}
  \country{Canada}
}

%%
%% By default, the full list of authors will be used in the page
%% headers. Often, this list is too long, and will overlap
%% other information printed in the page headers. This command allows
%% the author to define a more concise list
%% of authors' names for this purpose.
%\renewcommand{\shortauthors}{Trovato and Tobin, et al.}

%%
%% The abstract is a short summary of the work to be presented in the
%% article.
\begin{abstract}

The purpose of scholarly peer review is to evaluate the quality of scientific manuscripts. However, study after study demonstrates that peer review neither effectively nor reliably assesses research quality. 
Empirical standards attempt to address this problem by modelling a scientific community's expectations for each kind of empirical study conducted in that community. This should enhance not only the quality of research but also the reliability and predictability of peer review, as scientists adopt the standards in both their researcher and reviewer roles. However, these improvements depend on the quality and adoption of the standards. This tutorial will therefore present the empirical standard for mining software repositories, both to communicate its contents and to get feedback from the attendees. The tutorial will be organized into three parts: (1) brief overview of the empirical standards project; (2) detailed presentation of the repository mining standard; (3) discussion and suggestions for improvement. 

\end{abstract}



%%
%% The code below is generated by the tool at http://dl.acm.org/ccs.cfm.
%% Please copy and paste the code instead of the example below.
%%

%%
%% Keywords. The author(s) should pick words that accurately describe
%% the work being presented. Separate the keywords with commas.

\maketitle

\section{Overview of Empirical Standards Project} \label{sec:overview}
% \todo{TODO for Preetha - Can you please elaborate this section using the hints given below}

The Empirical Standards project aims to improve review quality, consistency, and predictability in the peer-review process for empirical studies by creating brief public documents that model our community's expectations for empirical research \cite{ralph2020empirical}. Creating and evolving these standards, and then reconstituting peer review processes around them, should also improve research quality and consensus. 

The project started with the ACM SIGSOFT special initiative to improve paper and peer review quality  \cite{ralph2020acm}. Two years and 50+ contributors later, Ralph \textit{et al.} \cite{ralph2021acm} released the standards and a prototype reviewing tool \cite{arshad2021towards}. The standards and review tools are available online.\footnote{\url{https://acmsigsoft.github.io/EmpiricalStandards/docs/}} The quickest way to grasp what is meant by an ``empirical standard'' is to go to the standards website and look at a standard for a familiar methodology. 

Critically, the standards are method-specific. The standard for experiments is very different from the standard for case studies. The standard for questionnaire surveys is very different from the standard for simulations. Standards must be method-specific to foster diversity in research and avoid cross-paradigm criticism (e.g. one should not criticize a case study for lack of generalizability because that's not what a case study is for \cite{stol2018abc}). However, the standards share a common format, including several sections: 
\begin{itemize}
    \item Application: how to determine whether this standard applies to a given manuscript
    \item Essential attributes: properties a manuscript must have to be acceptable in any peer reviewed venue
    \item Desirable attributes: properties that may enhance the rigor and quality of a manuscript but are not always necessary
    \item Extraordinary attributes: properties associated with award-quality research
    \item Anti-patterns: common problems seen in this kind of study
    \item Invalid criticisms: critiques reviewers should not make about this kind of study
    \item Suggested reading: references to helpful works about the methodology
    \item Exemplars: published manuscripts that effectively demonstrate some (not necessarily all) of the essential, desirable or extraordinary attributes.
\end{itemize}

Each standard was initially developed by a small team of scientists experienced in that method. Much care was taken to craft the essential attributes, as these will determine whether a manuscript is accepted for publication.  

However, research methods constantly evolve along with associated expectations for them; 
hence, this project aims to constantly update the standards to foster and incorporate emerging expectations. An empirical standard is supposed to \textit{model}, not \textit{set}, a community's expectations around rigor. We therefore encourage interested readers to suggest improvements to the standard by raising a pull-request on the GitHub repository.\footnote{\url{https://github.com/acmsigsoft/EmpiricalStandards}} and raising a pull-request.

\section{Motivation and Objectives}

To improve peer review and paper quality, the standards must be adopted in several ways: researchers using the standards to design studies and prepare manuscripts; reviewers using the standards to evaluate manuscripts; editors and program chairs using the standards to define quality for their venues. This motivates the twin aims of our tutorial:

\begin{enumerate}
    \item to communicate the contents of the repository mining standard to attendees and help attendees understand how to apply the standards in their various roles;
    \item to hear the attendee's feedback on the repository mining standard, understand their challenges and concerns (if any), and conceive potential improvements to the standard.
\end{enumerate}

\section{Tutorial Format}

The tutorial comprises three parts. It will begin with a brief overview of the empirical standards project.
% covering approximately the same territory as Section \ref{sec:overview}. We will then explain the repository mining standard, and lead a discussion about its current usefulness and potential improvements.
During the overview, we will explain the motivation for and goals of the empirical standards, as well as how the repository mining standard was developed, and the process for updating and improving the standards. 

In the second part of the tutorial,
we will explain how to interpret and use the repository mining standard. We will explain its main elements, as follows. 

\begin{description}
\item [Application:] 
The standard applies to software engineering studies that use automated techniques to extract data from large-scale data repositories and quantitatively analyze the contents mined from the repositories.

\item [Essential Attributes:]
The standard specifies the minimum essential attributes that the study must explain.
These attributes include the description and justification of data sources, repository selection criteria, and the procedure of data extraction.

\item [Desirable and Extraordinary Attributes:]
The standard summarizes attributes that are not universally necessary but tend to improve the quality of this kind of study. 
These attributes include aspects related to supplementary material, hypothesis testing, qualitative analysis of construct validity and dataset quality.
The standard also discusses extraordinary attributes such as establishing causality among the studied variables.

\item [Anti-patterns:]
The standard discusses several anti-patterns that a repository mining study must avoid, including limiting analysis to quantitative description, convenience sampling without good selection criteria, and
presenting insufficient details about the data processing steps.

\item [Invalid Criticisms:]
The standard provides a list of common but invalid criticisms including
needing to include more repositories
and expecting different sources of repositories or data than those selected and justified in the study. Reviewers should abstain from lodging such criticisms. 

\item [Suggested Reading:]
The standard lists a set of articles from the community that provide comprehensive treatment to one or more aspects included in the standard.

\item [Exemplars:]
The standard include references to some good example of software repository mining studies.
\end{description}

\balance

% The third part of the tutorial will be a discussion about its current usefulness and potential improvements.
In the third part of the tutorial,
the presenters will open the floor to the attendees to provide feedback on the current standard.
Additionally, suggestions will be sought to improve its usefulness and adoption by the research community. We hope that up to half of the session can be dedicated to discussion and feedback.

%\section{Target Audience}

\section{Speaker Biographies}
\noindent \textbf{Preetha Chatterjee}, \textit{Ph.D. (University of Delaware),} is an Assistant Professor at Drexel University. Her research interests are in improving software engineers’ tools and environments through empirical data analysis, natural language processing and machine learning.  She serves on the OC/PC for several conferences such as ICSE, MSR, ICSME, and SANER. 
\smallskip

\noindent \textbf{Tushar Sharma}, \textit{PhD (AUEB, Greece), MS (IIT-Madras, India),} is an assistant professor at Dalhousie University. His research interests include software quality, refactoring, and applied machine learning for software engineering. He worked with Siemens Research for more than nine years. He co-authored Refactoring for Software Design Smells: Managing Technical Debt and two Oracle Java certification books. He has founded and developed Designite which is a software design quality assessment tool used by many practitioners and researchers worldwide. He is an IEEE Senior Member.
\smallskip

\noindent \textbf{Paul Ralph}, \textit{PhD (British Columbia),} is an award-winning scientist, author, consultant, and Professor of Software Engineering at Dalhousie University. His research intersects software engineering, human-computer interaction, and project management. Paul is the editor of the Software Engineering Empirical Standards.

%\section{Acknowledgments}
%
%Identification of funding sources and other support, and thanks to
%individuals and groups that assisted in the research and the
%preparation of the work should be included in an acknowledgment
%section, which is placed just before the reference section in your
%document.
%
%This section has a special environment:
%\begin{verbatim}
%  \begin{acks}
%  ...
%  \end{acks}
%\end{verbatim}
%so that the information contained therein can be more easily collected
%during the article metadata extraction phase, and to ensure
%consistency in the spelling of the section heading.
%
%Authors should not prepare this section as a numbered or unnumbered {\verb|\section|}; please use the ``{\verb|acks|}'' environment.

%%
%% The acknowledgments section is defined using the "acks" environment
%% (and NOT an unnumbered section). This ensures the proper
%% identification of the section in the article metadata, and the
%% consistent spelling of the heading.

\begin{acks}
We would like to acknowledge the amazing support of everyone who contributed to the Empirical Standards for Software Engineering Research. A complete list of contributors is available at: \newline \url{https://acmsigsoft.github.io/EmpiricalStandards/people/}
\end{acks}

%%
%% The next two lines define the bibliography style to be used, and
%% the bibliography file.
\bibliographystyle{ACM-Reference-Format}
\bibliography{acmart.bib}

\end{document}